\documentclass[twocolumn,showpacs,preprintnumbers,amsmath,amssymb,floatfix]{revtex4}
\input{epsf}

\usepackage{graphicx}
\usepackage{dcolumn}
\usepackage{bm}

\begin{document}

\preprint{APS/123-QED}

\title{Dynamical stability of entanglement between spin ensembles}
\author{H. T. Ng and S. Bose}
\affiliation{{Department of Physics and Astronomy, University
College London, Gower Street, London WC1E 6BT, United Kingdom}}
\date{\today}

\begin{abstract}
We study the dynamical stability of the entanglement between the two spin ensembles in the presence of
an environment.  For a comparative study, we consider the two cases: a single spin ensemble, and two ensembles linearly
coupled to a bath, respectively.
In both circumstances, we assume the validity of the Markovian approximation for the bath.
We examine the robustness of the state by means of the growth of the linear entropy
which gives a measure of the purity of the system.  We find out macroscopic entangled
states of two spin ensembles can stably exist in a common bath.
This result may be very useful to generate and detect macroscopic
entanglement in a common noisy environment and even a stable macroscopic memory.
\end{abstract}


\maketitle

\section{Introduction}
Quantum entanglement is a fundamental concept in quantum mechanics.
It gives rise to Einstein-Podolsky-Rosen (EPR) paradox \cite{Einstein} and violates
a generalization of Bell's inequality \cite{Popescu}.  It is
also the physical ingredient of quantum information processing (QIP) such
as quantum communication (including quantum teleportation \cite{Bennett1} and dense coding \cite{Bennett2},
etc.).  To perform quantum communication, it is required to generate entangled
states between two distant locations \cite{Ekert}.  Recently, the entanglement has been
generated between two separated atomic ensembles \cite{Julsgaard1}.  In addition, quantum interfaces
between light and atoms have been experimentally
shown \cite{Julsgaard2} in which the state
of light was mapped onto collective excitations in atomic ensembles.
More recently, quantum light has been demonstrated to be storable in solid-state atomic medium \cite{Riedmatten}.
This may pave the way to implement ``long-distance'' quantum communication \cite{Duan}.

Atomic ensembles and solid-state atomic medium, can be
described as an ensemble of spin-half particles \cite{Duan,Riedmatten},
and inevitably suffer from decoherence due to coupling
to an external environment \cite{Breuer,Joos}.  This may heavily hinder the entanglement
generation and thus affects the performance of quantum memory \cite{Duan} using spin ensembles.
It is very crucial to study the robustness of the entangled states of spin
ensembles under decoherence effects.

\begin{figure}[ht]
\includegraphics[height=3.8cm]{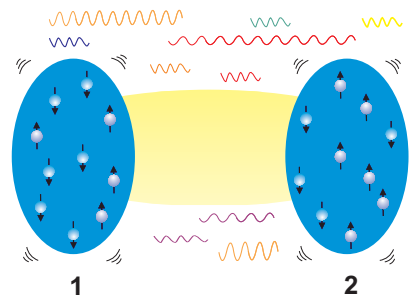}
\caption{ \label{fig1} (Color online) Two ensembles of spin-half
particles are coupled to a common bath.}
\end{figure}

In this paper, we study the spin ensembles linearly coupled to an independent
bath and the two spin ensembles coupled to a common bath respectively.
The two spin ensembles can be regarded as independently interacting with the bath
if they are well separated.  In contrast,  the two spin ensembles are effectively coupled to
the same bath if the separation between the two ensembles is much shorter than the correlation length
of the bath.  The schematic is shown in Fig. \ref{fig1}.
Here we assume that the Markovian approximation for the bath is valid in both situations.
In fact, the quantum behavior in coupling to independent baths is dramatically different to
coupling to common baths (collective decoherence) \cite{Massimo,Duan1}.
It is very important to examine the essential different features 
of quantum entanglement in these two decoherence models.  This leads to better
understanding of entanglement under the decoherence and inspires us to invent useful
methods to preserve the entanglement.

Here we are concerned about the sensitivity of the entangled states in the presence of
environment.  The robustness of the states can be quantified by how long the purity
of the spin ensembles can be maintained.  This can be measured by means of the growth of
linear entropy \cite{Joos}.  In the model of independent bath coupling, the rate of losing
the purity of maximally entangled systems are found to scale with the square of
the ``amount'' of entanglement.

However, we found that the entangled states of the two ensembles can exist
robustly in a common bath.  In particular, the singlet
states can even form a decoherence-free subspace (DFS) \cite{Lidar}.  This result shows that macroscopic
entanglement can persistently exist in a common noisy environment.  It motivates the further studies
of the macroscopic entanglement formation in the physical systems with the common bath.
This may be useful for QIP in atom-chip based \cite{Treutlein} and solid-state \cite{Makhlin}
systems which are required to perform short-ranged quantum communication \cite{Bose}.
For example, the two atomic Bose-Einstein condensates can be
coupled to the phonon modes of an elongated condensate \cite{Recati} to mediate the
entanglement between the two condensates.

\section{Independent Bath Model}
We study a decoherence model in which a spin ensemble is linearly coupled
to an environment.  In general, the Hamiltonian of the total system
can be written as \cite{Breuer}
\begin{eqnarray}
H&=&H_0+H_B+H_I,
\end{eqnarray}
where $H_0$ and $H_B$ are the Hamiltonian for the system and the bath respectively, and
$H_I$ is the interaction between the system and the bath.
We represent the spin ensemble in terms of the angular momentum operators:
$J=(J_{x},J_{y},J_{z})$.  We choose the
quantization axis in the $z-$direction such that
$J_{z}|m\rangle=m|m\rangle$ and
$J^2|m\rangle=j(j+1)|m\rangle$, where $j=N/2$ and $N$ is the number of
spin-half particles.   Here we consider the subspace for $j=N/2$ only because the
atomic states are totally symmetrized for a collection of identical spin-half particles \cite{Arecchi}.
The interaction Hamiltonian $H_I$ can be expressed in a general form as
\cite{Breuer}:
\begin{eqnarray}
H_I&=&\sum_{\alpha}J_{\alpha}\otimes{B}_\alpha,
\end{eqnarray}
where $B_{\alpha}$ is the bath operator and $\alpha=x,y$ and $z$.

We consider the interaction between the system and the environment
in the weak coupling regime.  We adopt the Markovian approximation
to write down the master equation of the system of the form \cite{Breuer}
\begin{equation}
\label{master1}
\dot{\rho}=-i[H_s,\rho]+\sum_{\alpha,\beta}\gamma_{\alpha\beta}[J_\beta{\rho}J_\alpha
-\frac{1}{2}\{J_{\alpha}J_\beta,\rho\}],
\end{equation}
where
$\gamma_{\alpha\beta}=\int^{\infty}_{-\infty}ds\langle{B_\alpha}(s)B_\beta(0)\rangle$ and $\alpha,\beta=x,y$ and $z$.

We can classify the decoherence model into three different cases according to the
couplings between the system and the bath. We define the one-axis model as exactly one axis
of the angular momentum system  coupled to the bath, say $z-$axis.  The one-axis model indeed gives a
realistic description in many quantum optical phenomena \cite{Breuer}.
The two-axis model can be defined as two axes coupled to the bath, however, we presume that
the coupling between the bath and one of the spin components is much stronger than the couplings of the remaining axes
to the bath.  The damping parameters can thus be written as $\gamma_{zz}\gg\gamma_{zx},\gamma_{xx}$.   
In the three-axis model, all axes are coupled to the bath.
We consider only one of axes are strongly coupled to the bath, i.e., $\gamma_{zz}\gg\gamma_{z\alpha},\gamma_{\alpha\beta}$
and $\alpha,\beta=x,y$.   The two- and three-axis models indeed
provide a more general scenario for the spin ensembles coupling to a bath.

We study the robustness of the states under the decoherence.
We can examine the stability of the
states based on the growth of linear entropy \cite{Joos}.  The linear entropy
provides a measure of the purity of a system \cite{Joos}.
The definition of the linear entropy is \cite{Joos}
\begin{eqnarray}
S_{\rm lin}=1-{\rm tr}(\rho^2).
\end{eqnarray}
A pure state gives a zero linear entropy $S_{\rm lin}=0$ and $0{\leq}S_{\rm lin}{\leq}1$.
Starting with a pure state, the rate of change of the linear entropy $\dot{S}_{\rm lin}$
is $-2{\rm tr}(\rho\dot{\rho})$ \cite{Joos}.

We take account of the entropy production of the early dynamics starting with a pure state.
This is sufficient to examine the sensitivity to the environment.
According to the master equation in Eq. (\ref{master1}), the rate of change of $S_{\rm lin}$
can be expressed in terms of the expectation values of angular momentum operators at the time $t=0$:
\begin{eqnarray}
\dot{S}_{\rm
lin}&=&2\sum_{\alpha,\beta}\gamma_{\alpha\beta}(\langle{J_\alpha}J_\beta\rangle-\langle{J_\alpha}\rangle\langle{J_\beta}\rangle).
\end{eqnarray}
Obviously, in the one-axis model,
the eigenstates of $J_z$ are formed and give a zero value of variance $(\Delta{J}_z)^2$.
Therefore, the Fock states $|m\rangle$, for $m=-j,-j+1,\ldots,j-1,j$, in the pointer basis \cite{Zurek} of $J_z$
naturally form the DFS in the one-axis model.

Now we study the robustness of the entangled states of two spin ensembles interacting with
the bath independently.  We consider a general entangled state of the form
\begin{eqnarray}
\label{ent1}
|\Psi_{\rm ent}\rangle=\sum^{\tilde{N}}_{m=-\tilde{N}}c_m|m,-m\rangle,
\end{eqnarray}
where $\tilde{N}={\rm min}\{j_1,j_2\}$ and $c_m$ is the probability coefficient.
The schemes for producing this entangled state in Bose-Einstein condensates has been proposed \cite{Dunningham}.
This state is useful for entanglement-based quantum communications with Bose-Einstein condensates.
The pure-state entanglement can be quantified by the von-Neumann entropy which is defined as
\begin{eqnarray}
E_F&=&-{\rm tr}(\rho_1\ln\rho_1),
\end{eqnarray}
where $\rho_1={\rm tr}_2(\rho)$ is the reduced density matrix of $\rho$.  Here the von-Neumann
entropy $E_F$ of the state in Eq. (\ref{ent1}) is $-\sum_m|c_m|^2\ln|c_m|^2$.
In the one-axis model, the rate of change of the linear entropy is
\begin{equation}
\dot{S}_{\rm lin}{\approx}2(\gamma_{zz}+\gamma_{zz}')\Big[{\sum_m}|c_{m}|^2m^2-\Big(\sum_m|c_m|^2{m}\Big)^2\Big],
\end{equation}
where $\gamma_{zz}$ and $\gamma_{zz}'$ are the two damping parameters for the two ensembles
respectively.  For a highly (nearly maximal) entangled state, the probability coefficient $|c_m|$ is equal (roughly equal) to
$1/\sqrt{2\tilde{N}+1}$.  This gives a value of the von-Neumann entropy with $\ln|2\tilde{N}+1|$.
In this case, we can estimate the growth of the entropy $\dot{S}_{\rm lin}$ which is roughly equal to $2(\gamma_{zz}+\gamma_{zz}')\tilde{N}^2/3$ \cite{remark1}.
Hence, the rate of the loss of purity scales with the square of the ``amount''
of the entanglement (the Schmidt number, i.e., $2\tilde{N}+1$ \cite{Nielsen}).  
This means that the purity of macroscopic entanglement vanishes quickly when the two spin ensembles interact with the bath independently.
We do not claim that the entanglement is lost completely as the purity is decreased.
Indeed, there is another measure of entanglement for mixed states \cite{Bennett3}. However, we can
expect that the entanglement of formation becomes very small if the state is highly ``mixed'' \cite{remark2}.

\section{Common Bath Model}
We consider the decoherence model of the two spin ensembles linearly coupling
to a common environment.
The total Hamiltonian reads
\begin{eqnarray}
H&=&H_0+H_B+H_I,
\end{eqnarray}
where $H_0$, $H_B$ and $H_I$ are the Hamiltonian for the system, the bath and 
the interaction between the system and the bath respectively.  
We represent the $i-$th spin ensemble in terms of the
usual angular momentum operators:
$J_{i}=(J_{{i}x},J_{{i}y},J_{{i}z})$, where $i=1,2$.  We have
$J_{iz}|m\rangle_i=m|m\rangle_i$ and
$J^2_i|m\rangle_i=j_i(j_i+1)|m\rangle_i$, where $j_i=N_i/2$.   

Without loss of generality, the
interaction Hamiltonian $H_I$ for coupling to a common bath can be written of a form
\begin{eqnarray}
H_I&=&\frac{1}{2}\sum_{\alpha}[{\lambda}J_{1\alpha}+(2-\lambda)J_{2\alpha}]\otimes{B}_\alpha,
\end{eqnarray}
where $\lambda\in[0,2]$ is a coupling parameter and
$B_{\alpha}$ is the bath operator and $\alpha=x,y$ and $z$.
The common bath model is identical to the independent bath model if we 
set $\lambda=0$ or 2.
We classify our model as the same way to the independent bath
model  such that the decoherence model is
classified as the one-, two- and three-axis models in the common bath.

We consider the interaction between the system and the environment
in the weakly coupling regime.  This enables us to adopt the Markovian approximation
to write down the master equation as
\begin{equation}
\dot{\rho}=-i[H_s,\rho]+\sum_{\alpha,\beta}\gamma_{\alpha\beta}[L_\beta{\rho}L_\alpha
-\frac{1}{2}\{L_{\alpha}L_\beta,\rho\}],
\end{equation}
where $L_\alpha=[{\lambda}J_{1\alpha}+{(2-\lambda)}J_{2\alpha}]/2$ is the composite angular momentum operator and
$\gamma_{\alpha\beta}=\int^{\infty}_{-\infty}ds\langle{B_\alpha}(s)B_\beta(0)\rangle$.

Similarly, we adopt the rate of change of $S_{\rm lin}$ to study the robustness of the states
in the common bath.
The rate of change of $S_{\rm lin}$ is given by
\begin{eqnarray}
\label{varL}
\dot{S}_{\rm
lin}&=&2\sum_{\alpha,\beta}\gamma_{\alpha\beta}(\langle{L_\alpha}L_\beta\rangle-\langle{L_\alpha}\rangle\langle{L_\beta}\rangle).
\end{eqnarray}
But the variances in Eq. (\ref{varL}) are expressed
in terms of the composite angular momentum operators $L_{\alpha}$ instead.

It is noted that the entangled state $|\Psi_{\rm ent}\rangle$ in Eq. (\ref{ent1})
was found to be very robust in the collective decoherence \cite{Duan1}.
We first examine the stability of the entangled state $|\Psi_{\rm ent}\rangle$ in the one-axis model,
the quantity $\dot{S}_{\rm lin}$ is given by
\begin{equation}
\dot{S}_{\rm lin}{\approx}2\gamma_{zz}(\lambda-1)^2\Big[{\sum_m}|c_{m}|^2m^2-\Big(\sum_m|c_m|^2{m}\Big)^2\Big].
\end{equation}
We can estimate that $\dot{S}_{\rm lin}$ is about $2\gamma_{zz}(\lambda-1)^2\tilde{N}^2/3$ for the highly
entangled state with $|c_m|{\approx}1/\sqrt{2\tilde{N}+1}$.  This result is consistent with the rate of the growth
of entropy in the independent bath model.  The losing rate of the purity also scales with the square of the
 ``amount'' of the entanglement for highly entangled states.  
 However, the rate of the growth of the linear entropy can be
dramatically reduced if the parameter $\lambda$ is close to one.   This is the essential
feature in the collective decoherence.

In the one-axis model, the decoherence can be completely quenched in the limit of
$\lambda$ approaching one.  However, the decoherence cannot be eliminated in
the two-axis model even $\lambda$ is set to be one.
We can minimize the decoherence rate in Eq. (\ref{varL}) if its
variance $(\Delta{L_x})^2$ can be kept to be very small for $\lambda=1$.  This means that
the number of particles in each ensembles are nearly the same, i.e.,
$j_1{\approx}j_2{\approx}{j}$.   The quantum variance $(\Delta{L_x})^2$
of the entangled state $|\Psi_{\rm ent}\rangle$ is given by
\begin{equation}
\label{varJx}
(\Delta{L_x})^2\approx\sum_{m}[|c_m|^2+{\rm Re}(c^*_{m-1}c_m)][j(j+1)-m^2].
\end{equation}
The quantum fluctuation can be minimized if the condition is
satisfied:
\begin{eqnarray}
\label{mincond} [|c_m|^2+{\rm Re}(c^*_{m-1}c_m)]\rightarrow{0}.
\end{eqnarray} For example, the quantum fluctuations are greatly reduced
if we can take $c_m\approx{(2{\tilde{N}})^{-1/2}}$
and ${\rm Re}(c^*_{m-1}c_m)<0$.   For the entangled state $|\Psi_{\rm ent}\rangle$, the variance
$(\Delta{L_y})^2$ has the same form as the variance $(\Delta{L_x})^2$ and also the
cross correlations $\langle{L_xL_y+L_yL_x}\rangle$ is zero.
Therefore, the entangled state is very stable even in the three-axis model
if the condition (\ref{mincond}) can be achieved.

Having discussed the decoherence properties of the general entangled state,
we study the decoherence of eigenstates of the composite angular momentum system in
the ideal collective decoherence ($\lambda=1$).
In fact, the eigenstates of the composite angular momentum system in the $z-$direction
are formed a DFS in the one-axis model.  The composite eigenstate $|L,M\rangle$ can be written as
\begin{eqnarray}
|L,M\rangle&=&\sum^{j_{1,2}}_{m_{1,2}=-j_{1,2}}C^{LM}_{j_1m_1j_2m_2}|m_1,m_2\rangle,
\end{eqnarray}
where $C^{LM}_{j_1m_1j_2m_2}=\langle{m_1,m_2}|L,M\rangle$ is the Clebsch-Gordon coefficient
and $L=j_1+j_2,j_1+j_2-1,\ldots,|j_1-j_2|$ and $M=m_1+m_2$.
The eigenstate $|L,M=0\rangle$ is clearly an entangled state for the two spin
ensembles.  The entangled pairs are formed with $m_1=-m_2=m$ with the
total population number $M=0$.

For the state $|L,M=0\rangle$, we evaluate the quantity $\dot{S}_{\rm lin}$
which gives $\gamma_{xx}L(L+1)$ and $(\gamma_{xx}+\gamma_{yy})L(L+1)$ in the
two- and three-axis couplings respectively.
We can see that the state $|L=0,M=0\rangle$ forms a DFS even in the three-axis linear model
for $\dot{S}_{\rm lin}=0$.
Indeed, this singlet state has been found to be decoherence-free \cite{Lidar2} because
it is totally symmetric to the environment.
This singlet state $|L=0,M=0\rangle$ gives out the maximal
entanglement with $E_F=-\ln|2j+1|$ for $c_m=(-1)^{j-m}$ and $m=-j,-j+1,\ldots,j-1,j$.
Besides, the violation of the Bell inequality
for this singlet state $|L=0,M=0\rangle$ has been discussed
\cite{Peres}.

Nevertheless, we point out that there are difficulties in producing the ideal singlet state
$|L=0,M=0\rangle$ in experiments.  This is because it is very difficult to ensure
the same number of particles in each ensemble.  However, it can be easily shown that the state
$|L,M=0\rangle$ is also very robust in the common bath for the low values of $L$.
This means that the states with $M=0$ are possible to be prepared if the number of particles in the two
ensembles are very close.  Rather than detecting the stable entangled state $|\Psi_{\rm ent}\rangle$ in Eq. (\ref{ent1})
in a common noisy environment, one can also use them as quantum memory if suitable encoding and decoding mechanisms
can be found.

\section{Discussion}
We have studied the robustness of states in the cases of a spin ensemble
and the two spin ensembles coupled to a bath respectively.
We have shown that the totally different features in losing the purity in these
two cases.
In the independent bath model, the decay rate of the purity of the maximally
entangled systems scales as the square of ``amount'' of the entanglement.
This result is useful for understanding the decoherence of the entanglement
for two well separated systems such as atomic ensembles.

On the contrary, the entanglement can be preserved much longer if the two spin
ensembles are coupled to the common bath.  It provides a ground to detect macroscopic
pure-state entanglement in a common noisy environment.  In the future,
we will study the formation of macroscopic entanglement in the physical systems.
This may be useful for entangling two spin ensembles to perform ``short-distance''
quantum state transmission \cite{Bose}. 

\begin{acknowledgments}
The work of H.T.N. is supported by the Quantum Information
Processing IRC (QIPIRC) (GR/S82176/01). S.B. also thanks the
Engineering and Physical Sciences Research Council (EPSRC) UK for an
Advanced Research Fellowship and the Royal Society and the Wolfson
Foundation.

\end{acknowledgments}


\begin{thebibliography}{99}
\bibitem{remark1}
The upper bound of $\dot{S}_{\rm lin}$ can be given out
as $4(\gamma_{zz}+\gamma_{zz}')\tilde{N}^3/3$ for a general
entangled state in Eq. (\ref{ent1}).
\bibitem{remark2}
The definition of entanglement of formation for the mixed states is \cite{Bennett3}
\begin{equation}
E_c(\rho)={\rm min}\sum_ip_iS(\rho^i_1)
\end{equation}
where $S(\rho_1)=-{\rm tr}{\rho_1}\ln{\rho_1}$ is the von Neumann entropy and
take the minimum of the state $\rho_{12}=\sum_ip_i|\psi_i\rangle\langle{\psi_i}|$
with the reduced density matrix $\rho^i_1={\rm tr}_2(|\psi_i\rangle\langle\psi_i|)$.
It is easy to show that the probability $p_i$ is of the order of $j^{-1}_i$ if the
state becomes highly mixed.  This means that the entropy $E_c(\rho)$ is also very
small in this case.
\end{thebibliography}
\end{document}